\documentstyle[aasms4,epsfig]{article}

\newcommand{\derp}[2]{\frac{\partial #1}{\partial #2}}
\newcommand{\dderp}[2]{\frac{\partial^2 #1}{\partial {#2}^2}}
\newcommand{\derd}[2]{\frac{{\rm d}  #1}{{\rm d}  #2}}
\def\ov{\over}
\def\l{\left}
\def\r{\right}
\def\be{\begin{equation}}
\def\ee{\end{equation}}

\begin{document} 
  
\title
{Gravitational waves from the collapse and bounce of a stellar core in
tensor-scalar gravity}  

\author{J\'er\^ome Novak\altaffilmark{1} and Jos\'e
M$^{\underline{\mbox{a}}}$.\,Ib\'a\~{n}ez} 
\affil{Departamento de Astronom\'\i a y
Astrof\'\i sica, Universidad de Valencia, 46100 Burjassot, Spain}

\altaffiltext{1}{also at D\'epartement d'Astrophysique Relativiste et
de Cosmologie 
  (UMR 8629 C.N.R.S.), Observatoire de Paris,
   Section de Meudon, F-92195 Meudon Cedex, France }

\begin{abstract} 
Tensor-scalar theory of gravity allows the generation of gravitational
waves from astrophysical sources, like Supernov\ae{}, even in the
spherical case. That motivated us to study the collapse of a
degenerate stellar core, within 
tensor-scalar gravity, leading to the formation of a neutron star
through a bounce 
and the formation of a shock. We discuss in this paper the effects of
the scalar field on the 
evolution of the system, as well as the appearance of strong
non-perturbative effects of this scalar field (the so-called
``spontaneous scalarization''). As a main result, we describe the resulting 
gravitational monopolar radiation (form and amplitude) and discuss the
possibility of its detection by the gravitational detectors currently
under construction, taking into account the existing constraints
on the scalar field. From the numerical point of view it is worthy to
point out that 
we have developed a combined code which uses pseudo-spectral methods, for 
the evolution of the scalar field, and High Resolution Shock-Capturing 
schemes, for the evolution of the hydrodynamical system. 
Although this code has been used to integrate the field equations of that 
theory of gravity, in the spherically symmetric case, a by-product of 
the present work is to gain experience for an ulterior extension
to multidimensional problems in Numerical Relativity of such 
numerical strategy. 


\end{abstract} 
 
\section{Introduction}\label{s:intro}
 
Tensor-scalar theories of gravitation (\cite{DEF92}) appear to be
interesting alternatives to Einstein's General Relativity, which
naturally arise from the low-energy limit of superstring theories
(\cite{DAP94}). These theories also allow for a
``graceful exit'' from inflation (\cite{GBQ90}). One of their main
observational predictions is the emission of {\em monopolar\/}
gravitational radiation, from quasi-spherical astrophysical
objects. Previous works on collapses in tensor-scalar theories dealt with the
gravitational collapse in tensor-scalar theories started with studying
the Oppenheimer-Snyder (dust) collapse (\cite{OS39}) in the frame of
the so-called ``Jordan-Fierz-Brans-Dicke'' theory, in which the
coupling function (\ref{e:cfunc}) depends only on one parameter
(\cite{SNN94} and \cite{SST95}). The next work by \cite{HCN97}
considered a two-parameter coupling function (as in this work), but
neglecting any non-perturbative effect from the scalar field. However,
these non-perturbative effects can be very important in neutron stars,
as have shown \cite{DEF93}, making an analogy between this
``spontaneous scalarization'' of neutron stars and the spontaneous
magnetization of ferromagnets (\cite{DEF96}). \cite{NOV98a} (paper I
hereafter) dealt with the
collapse of a neutron star to a black hole, including pressure and
non-perturbative effects. We here study the emission of monopolar
gravitational waves coming from the formation of a neutron star. The
problem is quite different since, according to Hawking's ``no-hair'' theorem
(\cite{HAW72}), a black hole cannot possess any scalar charge, neutron
stars appear to be the astrophysical object with the strongest
possible scalar charge. Their formation (during a Supernova type Ib or
type II event) is therefore a powerful and interesting source of
gravitational radiation.
The purpose of this paper is to present results from the
numerical study of the collapse of a stellar core which forms a
neutron star, within the framework of tensor-scalar theories and,
namely, to
explicit the resulting monopolar gravitational radiation.

Following paper I, we will consider only one scalar
field $\varphi$ and we will use two conformally
related metrics $\tilde{g}_{\mu \nu}= a^2(\varphi)g^*_{\mu\nu}$,
called the {\it Jordan-Fierz} (or physical) metric
($\tilde{g}_{\mu\nu}$) and the {\it Einstein} metric
($g^*_{\mu\nu}$). All the quantities with a tilde are related to the
Jordan-Fierz metric, to which the matter is coupled. That means that
all non-gravitational experiments measure this metric, although the
field equations of the theory are better formulated in the Einstein
one. The tensor-scalar field equations are
\be
  R^*_{\mu \nu} - {1\over 2}g^*_{\mu \nu}R^*  = 2\partial_\mu \varphi
\partial_\nu \varphi - g^*_{\mu \nu}g_*^{\rho \sigma} \partial_\rho
\varphi \partial_\sigma \varphi + {8\pi G_*\over c^4} T^*_{\mu \nu}, 
\label{e:tscal}
\ee
for the tensor part, and
\be
 g^{\mu\nu}_* \nabla^*_\mu \nabla^*_\nu \varphi = -4\pi G_*
\alpha(\varphi)T_* \label{e:ondsc}
\ee
for the scalar field.

The Einstein-frame stress-energy tensor $T_*^{\mu\nu}$ is related to
the physical one by the arbitrary coupling function $a(\varphi)$
\be
T^\mu_{*\nu} = a^4(\varphi) \tilde{T}_\nu^\mu .
\ee 
The function $\alpha(\varphi)$ is the logarithmic derivative of
$a(\varphi)$ and represents the field-dependent coupling strength
between matter and the scalar field. We will assume it is well
represented by a linear function (see \cite{NOV98b} for more
details):
\be
\alpha(\varphi) = \alpha_0 + \beta_0 \times(\varphi - \varphi_0).
\label{e:cfunc}
\ee
The arbitrary function $a(\varphi)$ is determined by the two
parameters $(\alpha_0,\beta_0 )$ ($\varphi_0$, the scalar field at
spatial infinity, is redundant with $\alpha_0$). We will also consider
the collapse to be spherically symmetric, since we are concerned with
monopolar radiation, which should be dominant over higher terms in
such events. 

The paper is organized as follows: Section~\ref{s:eqnum} presents the
gauge choice, the tensor-scalar equations derived in this gauge
(Sec.~\ref{ss:eqrgps}) and the numerical method used for integrating
them (Sec.~\ref{ss:intnum}), as well as the performed tests
(Sec.~\ref{ss:tests}). Section~\ref{s:simu} is dedicated to the
simulations which were done; namely, the physical models used
(\ref{ss:mode}), results of the simulations (\ref{ss:resu}) and a
discussion on the resulting gravitational waves and their interaction
with the detectors (\ref{ss:gwave}). Finally, Section~\ref{s:conc}
summarizes the results and gives some concluding remarks.
  
\section{Equations and numerical issues} \label{s:eqnum}

As in Paper I, we have chosen the {\em Radial Gauge} and {\em
Polar Slicing} (Schwarzschild-like coordinates), the metric being
diagonal:
\be
ds^2 = -N^2(r,t)dt^2 + A^2(r,t)dr^2 + r^2(d\theta^2 +
 \sin^2\theta d\phi^2), \label{e:metric}
\ee
where $N(r,t)$ is called the {\em lapse function}. The metric 
$g^*_{\mu\nu}$ will
often be described by the three auxiliary functions $\nu(r,t)$, $m(r,t)$ and
 $\zeta(r,t)$ defined by
\[
N(r,t) = \exp(\nu(r,t)), \]
\[
A(r,t) = \left(1-{2m(r,t)\over r}\right)^{-1/2}, \]
\[
\zeta(r,t) = \ln \l({N\over A}\r).
\]
All coordinates are expressed in the Einstein frame, and asterisks are
 omitted. However, ``physical'' quantities will often be written with
a tilde, that is in the Jordan-Fierz frame. The matter is supposed to
be modeled by a perfect fluid:
\be
\tilde{T}_{\mu\nu} = (\tilde{e}+\tilde{p})\tilde{u}_\mu \tilde{u}_\nu 
+ \tilde{p} \tilde{g}_{\mu\nu}, \label{e:flparf}
\ee
where $\tilde{u}_\mu$ is the 4-velocity of the fluid, $\tilde{e}$ is the total 
energy density
(including rest mass) in the fluid frame and $\tilde{p}$ is the pressure. 

\subsection{Equations in RGPS gauge} \label{ss:eqrgps}

As far as the hydrodynamics are concerned, we will use a set of {\em
primitive variables} $\{\tilde{n}_B,\tilde{e}, U \}$, which are
respectively the baryonic and the total energy densities in the fluid
frame, and the fluid radial velocity:
\[
U={A\over N}{\tilde{u}^r \over \tilde{u}^0};
\]
$\Gamma = (1-U^2)^{-1/2}$ being the Lorentz factor of the fluid. We
also use the variable:
\[
\tilde{E} = - \tilde{T}^0_0 =\Gamma^2(\tilde{e}+\tilde{p}) -
\tilde{p},  
\]
and three ``scalar-field'' variables:
\[
\eta = {1\over A}{\partial \varphi \over \partial r},\  
\psi = {1\over N}{\partial \varphi \over \partial t},\  
\Xi = \eta ^2 + \psi ^2 .
\]

Considering the conservation of the stress-energy tensor and of the
baryonic number (Cf.~Paper I), one can write the hydrodynamical
equations for evolution in a hyperbolic form\footnote{the equation for
the density is not strictly in that form, see hereafter, at the end of
the section}:
\be
\derp{\vec{u}}{t} + {1\ov r^2}\derp{}{r}\l[ r^2{N \ov
A}\vec{f}(\vec{u}) \r] = \vec{s}(\vec{u}), \label{e:conshyp}
\ee
where $\vec{u}=\{D^*,\mu^*,\tau^*\}$ is the {\em vector of unknowns}
(evolved quantities), defined as
\begin{eqnarray}
D^* &=& a^4(\varphi) A \Gamma \tilde{n}_B, \nonumber \\
\mu^* &=& a^4(\varphi) \l( \tilde{E} + \tilde{p} \r) U,
\label{e:defhyp} \\
\tau^* &=& a^4(\varphi) \tilde{E} - D^*. \nonumber
\end{eqnarray}
The associated vectors of fluxes $\vec{f}(\vec{u})$ and sources
 $\vec{s}(\vec{u})$ being (we define $q_\pi=4\pi G_*/c^4$):
\begin{eqnarray}
f_{D^*} &=& U D^* \nonumber \\
f_{\mu^*} &=& \mu^*U + a^4(\varphi)\tilde{p} \label{e:deflux} \\
f_{\tau^*} &=& \mu^* - UD^*. \nonumber
\end{eqnarray}
and
\begin{eqnarray}
s_{D^*} & = & \alpha (\varphi )a(\varphi )D^*  {\partial \varphi \over
\partial t} \nonumber
\\
s_{\mu^*} & = & NA \l\{ \l( \mu^* U - \tau^* -D^* \r) \l( 2q_\pi rp^*
+{G_*m\ov r^2 c^2} + {\alpha(\varphi)\eta  \ov A} \r) + p^* {G_* m \ov
r^2 c^2} + {2p^* \ov A^2 r} \right. \nonumber \\
&&\left. + {3 \alpha(\varphi) p^* \eta \ov A}
-2r\mu^* \eta \psi - {r\ov 2}\Xi \l( \tau^*+D^* + p^* \r) (1+U^2) \r\}
\label{e:defsou} \\
s_{\tau^*} & = & \l( \tau^* + D^* + p^* \r) NAr \l\{ (1+U^2)\psi \eta
+ U \Xi \r\} - s_{D^*}. \nonumber
\end{eqnarray}
With this formulation, we easily recover the set of hydrodynamical
equations in General Relativity, in the limit ($\varphi \to \varphi_0,
\ a(\varphi) \to 1$), as described in \cite{RIM96}. Here, however, the
evolution equation for the relativistic density ($D^*$) is not
strictly in the form (\ref{e:conshyp}), but writes:
\be
\derp{D^*}{t} + {a(\varphi)\ov r^2}\derp{}{r}\l[ r^2{N \ov
A}f_D^* \r] = s_{D^*}, \label{e:evodens}
\ee

The system of equations is completed by an equation of state
$\tilde{p}=\tilde{p}(\tilde{n}_B,\tilde{e})$, which will be detailed
in Sec.~\ref{ss:mode}, and the field equations for the metric and the
scalar field:
\begin{eqnarray}
\derp{m}{r} &=& {c^2 \ov G_*} r^2 \l(\Xi+q_\pi(\tau^*+D^*) \r),
\label{e:dmdr} \\
\derp{\nu}{r} &=& {q_\pi A^2 \ov 2} \l({mc^2 \ov 4\pi r^2} + r(\Xi +
f_{\mu^*}) \r), \label{e:dnudr} \\
\dderp{\varphi}{t} &=& e^{2\zeta}\l( \Delta \varphi + \derp{\zeta}{r}
\derp{\varphi}{r} \r) + \derp{\zeta}{t} \derp{\varphi}{t} - {q_\pi
\ov 2} \alpha(\varphi) T_* \label{e:wavephi}.
\end{eqnarray}
For more details on the derivations of this system, see \cite{GOU91},
Paper I and \cite{RIM96}. The key point is the choice of the
conserved quantities (\ref{e:defhyp}) which, in tensor-scalar
theories, keeps the system in a strictly hyperbolic form
(\ref{e:conshyp}). 

\subsection{Numerical integration} \label{ss:intnum}

As it will be discussed in Sec.~\ref{ss:mode}, shocks are expected to
appear in the simulations. It is then clear that the
numerical integration must be able to handle such discontinuities. The
code used in Paper I is based on pseudo-spectral techniques and so, is
not well adapted for this kind of problem. However, in one dimension
\cite{BOM91} have adapted the spectral methods to the case where a
shock was present. This technique will not be used here, since we are
interested in generalizing (in a future work) techniques developed
here to two- and three-dimensional calculations. All the details of the
numerical techniques will not be described here, one can refer to
\cite{RIM96} and \cite{MAR97} for the Godunov-type (High-Resolution
Shock-Capturing, hereafter HRSC) methods and to Paper I and references
therein for the spectral methods.

The idea is then to use two different numerical grids to describe the
same physical space (using the same coordinates). These two grids
allow for two different numerical techniques for solving the 
(\ref{e:conshyp}) --
(\ref{e:wavephi}), governing the evolution of the
matter and fields. The HRSC methods, on a grid with a rather large
number of points (typically a few hundreds), are used to solve the
hydrodynamical hyperbolic system (\ref{e:conshyp}), for they are the
most efficient in handling discontinuous quantities, such as density
or velocity after the bounce. Spectral methods
are used to describe gravitational fields ($g^*_{\mu \nu}$ and
$\varphi$), which are smooth, as well as their first derivative. In this work,
for the equations of the Sec.~\ref{ss:eqrgps} only the wave equation
(\ref{e:wavephi}) has been integrated by those means, the two
constraint equations, (\ref{e:dmdr}) and (\ref{e:dnudr}), giving the
metric are very simple in the RGPS gauge and were integrated by usual
finite-difference method. 

Last but not least, the numerical information of the various fields
has to be passed from one grid to the other paying special attention
to minimize the numerical noise, not to destabilize the integration
schemes. Getting 
from the spectral grid (spectral representation of functions as a
truncated series of Chebyshev polynomials) to the HRSC one shows no
difficulties, since a function, represented by its
Chebyshev coefficients, can be evaluated at any point of its
definition interval with a high precision as a (truncated) series and
using the properties of Chebyshev polynomials (see
e.g. \cite{BGM97}). Nevertheless, the interpolation in the other way
(from HRSC grid to the spectral one) has to be ``smooth'' enough in
order not to bring high frequency terms. The method used here meets
these requirements, since functions are interpolated in the smoothest
way, introducing no spurious information (\cite{BON98}). The
interpolated points are such that the resulting amplitude of the
function $f$, calculated as 
$$
\int_{\rm interval} \l( \frac{d^2f}{dx^2} \r)^2,
$$
is minimized. This ensures that the Chebyshev coefficients of the
interpolated function decrease fast enough and the overall series
converges.

The resulting code (RSSM in the next,RS stands for Riemann
Solvers and SM for Spectral Methods) is then just the
combination of an HRSC module, integrating in time the hydrodynamical
variables $\{\tilde{n}_B,\tilde{e}, U \}$ and resolving the constraint
equations giving $m$ and $\nu$. The source term of the wave equation
(\ref{e:wavephi}) is then evaluated on the HRSC grid and interpolated
to the spectral grid. Then, the wave equation is integrated to get
the scalar field, which is computed, on the HRSC grid from its
Chebyshev coefficients. The typical number of points used for the HRSC
grid is between 300 and 500; for the spectral grid it is between 65
and 129. 

\subsection{Tests of the code} \label{ss:tests}

The {\it HRSC part} of the code comes directly from the code presented in
\cite{RIM96}, in which numerous tests are shown. The {\it spectral part},
solving the wave equation, has been presented and tested in Paper
I. The {\it interpolation procedure} has been tested on various analytical
functions. The combined
code, as a whole, has been checked by setting the scalar field to its
asymptotic value $\varphi_0$
(as well as the coupling function $\alpha(\varphi)$ (\ref{e:cfunc})),
which gave us the 
general-relativistic results. On the other hand, putting the pressure
to zero enabled us to retrieve the {\it dust collapse} of previous
studies (\cite{SNN94}, \cite{SST95} and \cite{HCN97}). 

\begin{figure}
\centerline{ \epsfig{figure=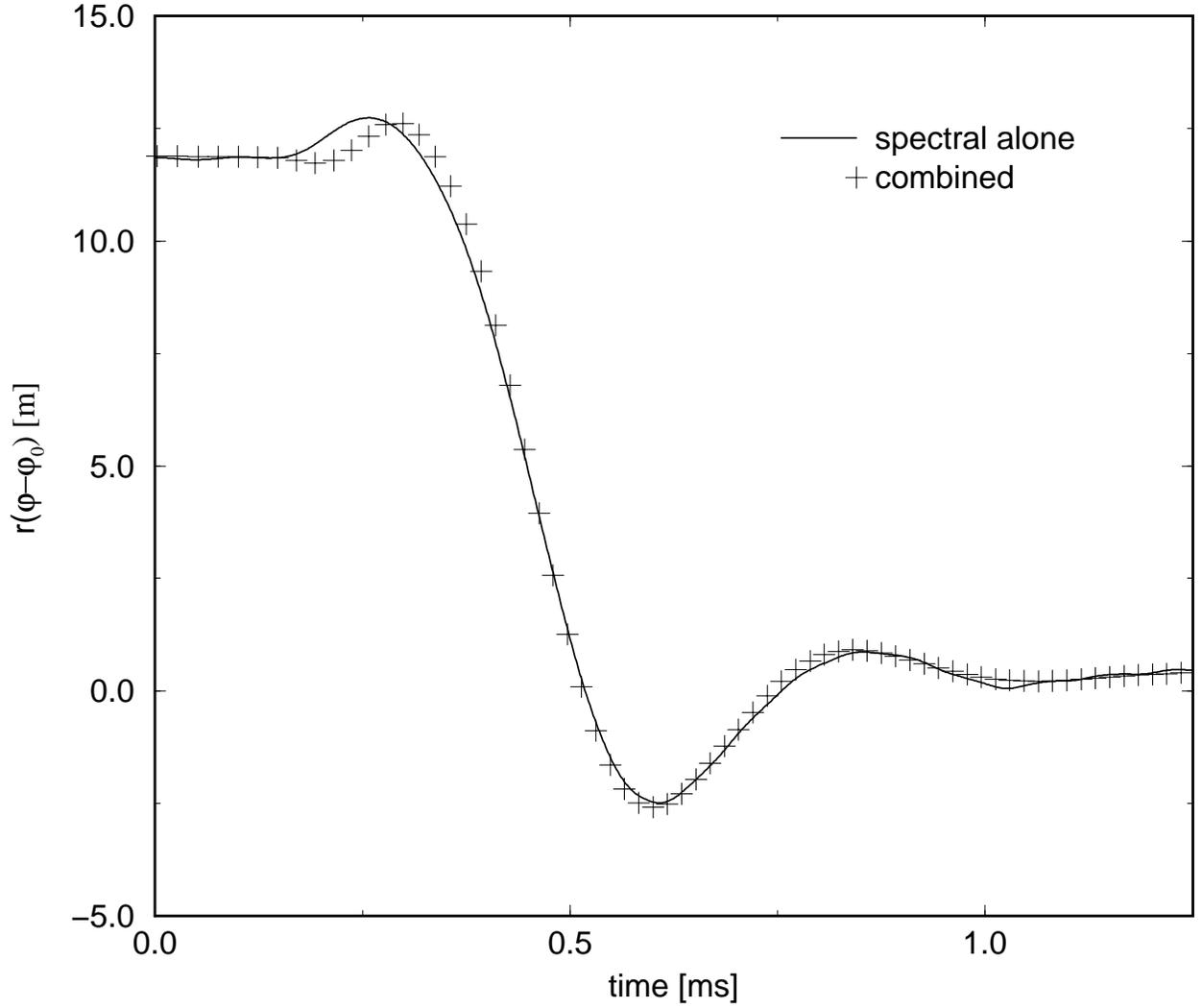,height=\hsize,angle=-90} }
\caption[] {Comparison between the pure spectral code and the combined
(HRSC and spectral) for the collapse of a  neutron
star, of baryonic mass $2.2\ \rm{M}_{\sun}$ and with $\alpha_0 = 2.5
\times 10^{-3}$, $\beta_0 = -4$. Time origin is arbitrary.}
\label{f:comf}
\end{figure}

Finally, we tried to follow the {\it collapse of a neutron star to a black
hole}, to compare with previous results from Paper I, but with this
combined code. The initial model 
taken is a neutron star with a polytropic equation of state (see
hereafter eq.~\ref{e:eospoly}). Since HRSC methods do not achieve the same
precision as spectral ones, the collapse has to be triggered in a
artificial way: decreasing the adiabatic index $\gamma$ or the $K$ coefficient
of the equation of state:
\begin{eqnarray}
\tilde{e}(\tilde{n}_B) = \tilde{n}_B \tilde{m}_B+K{\tilde{n}_0\tilde{m}_B
\ov \gamma -1}\l({\tilde{n_B} \ov \tilde{n}_0}\r)^\gamma \nonumber \\
\tilde{p} = K \tilde{n}_0\tilde{m}_B\l({\tilde{n_B} \ov \tilde{n}_0}
\r)^\gamma. \label{e:eospoly}
\end{eqnarray}
$\tilde{m}_B$ is the mass of one baryon ($=1.66 \times 10^{-27} \rm{
kg}$) and $\tilde{n}_0 = 0.1 \rm{fm}^{-3}$. The resulting scalar
waveform (see Sec.~\ref{ss:resu}), computed by both codes separately
is displayed 
on Fig.~\ref{f:comf}. The discrepancy is quite important at the
beginning of the collapse, where the instability has to develop and
the HRSC methods are less accurate. In addition, the boundary
conditions imposed on the hydrodynamical system are not rigorously the
same: as mentioned in Paper I, the spectral code uses a comoving
grid, and there is only one boundary condition to impose which is
chosen to be
$
\l.{1\ov r^2}{\partial\ov \partial r} r^2 U\r|_{r=R_{star}} = 0;
$
the HRSC methods use an \emph{Eulerian} grid and impose three boundary
conditions, chosen to be $f_{D^*} = f_{\mu^*} = f_{\tau^*} = 0$ at the
boundary of the grid. This can influence the beginning of the collapse,
but after that, the evolution is fast enough and such discrepancies
disappear. 

\section{Simulations} \label{s:simu}

As it has been mentioned above, the physical scenario we are
considering is that of the formation of a neutron star.

\subsection{The model} \label{ss:mode}

One can think of two ways to come to a neutron star: the collapse of
the iron core of a massive star at the end of its evolution, i.e. the
standard mechanism of type II supernova (\cite{ST83}), or the
accretion induced collapse of a white dwarf which, under some
conditions, can be non-explosive (\cite{CAN90}). In any case, the
degeneracy pressure of the electrons can no longer support the
gravity of the core and the collapse is unavoidable. When the central
density becomes comparable to that of nuclear matter, the strong
interactions between nucleons stiffen the equation of state. A strong
shock is then generated according to the well known model for {\it
prompt} mechanism of supernova (see \cite{MUL97} for details).

Since we are interested in the emission of monopolar gravitational
radiation in the framework of a tensor-scalar theory of gravity, we
have simplified the micro-physics concerning the equation of state (see
below) and the neutrino transport (which has been neglected). The
equation of state we consider is that of a perfect fluid
for which the adiabatic index
\begin{equation}
\gamma = \left. \derd{\ln \tilde{p}}{\ln \tilde{n}_B}
\right|_{\rm{adiab.}} \label{e:indadiab} 
\end{equation}
is not constant throughout the evolution of the collapsing star, but
is {\it a priori} dependent on the density, temperature and
chemical composition of matter. During the collapse, the adiabatic
index is supposed to be lower than $4/3$, but when the strong
interactions between nucleons become dominant over other pressures, the
adiabatic index grows rapidly above $4/3$ and even $2$. We
can also consider that all the matter in the core follows roughly the
same trajectory in the density-temperature space (see \cite{ARN77} and
\cite{VRI78}). In a first approximation, the adiabatic index thus depends
only on the density:
\begin{equation}
\gamma(\tilde{n}_B) = \gamma_{\rm{min}} + S\l(
\log_{10}(\tilde{\rho}) - \log_{10}(\rho_{\rm{bounce}}) \r)
\label{e:gamaro}
\end{equation}
where $\tilde{\rho}= \tilde{m}_B \tilde{n}_B$, and
($\rho_{\rm{bounce}}$, $S$) are two parameters of the equation of
state. There are other simplified versions of the equation of state,
which try to minimize  the complex physics involved (see
e.g. \cite{MUL97}). In this work we use (\ref{e:gamaro})and we have
taken $\gamma_{\rm{min}}=1.33$ in all our runs. 
\begin{table}
\begin{center}
\begin{tabular}{||c|c|c|c|c|c||}
\hline
model & A & B & C & D & E \\
\hline
 $\alpha_0$ & 0. & 0. & -0.01 & -0.01 & $-1\times 10^{-3}$ \\
 $\beta_0$ & 0 & 0 & -4 & -4 & -7 \\
 $\tilde{\rho}_{\rm{bounce}}$ [$\rho_{\rm{nuc}}$] & 1.5 & 15 & 1.5 &
15 & 1.5 \\
 $S$ & 1 & 5 & 1 & 5 & 1 \\
\hline
\end{tabular}
\caption{\label{t:efen}
Parameters of the equation of state and of the coupling function
(\ref{e:cfunc}), for the runs presented in this work.
We took 1 $\rho_{\rm{nuc}}=1.66\times 10^{17} \ \rm{ kg \cdot
m}^{-3}$.}
\end{center}
\end{table}
Table~\ref{t:efen} summarizes the different parameter used for the
models studied in this work. Models A, C and E use ``standard''
values for the equation of state parameters, whereas models B and D
will produce a ``stiffer'' bounce and allow to explore higher central
densities. This second type of model, even if it seems less realistic,
allows for more relativistic regimes (the resulting neutron star being
more compact), with stronger shocks. The initial state is the
same as in \cite{RIM96}: a white dwarf of $1.39 M_{\sun}$, which is
the maximal mass for such objects, as derived in \cite{IBA84}. Models
A and B are the same as in \cite{RIM96}, and will be used as
general-relativistic calibrations, to compare with tensor-scalar
results. 

\subsection{Results} \label{ss:resu}

\begin{figure}
\centerline{ \epsfig{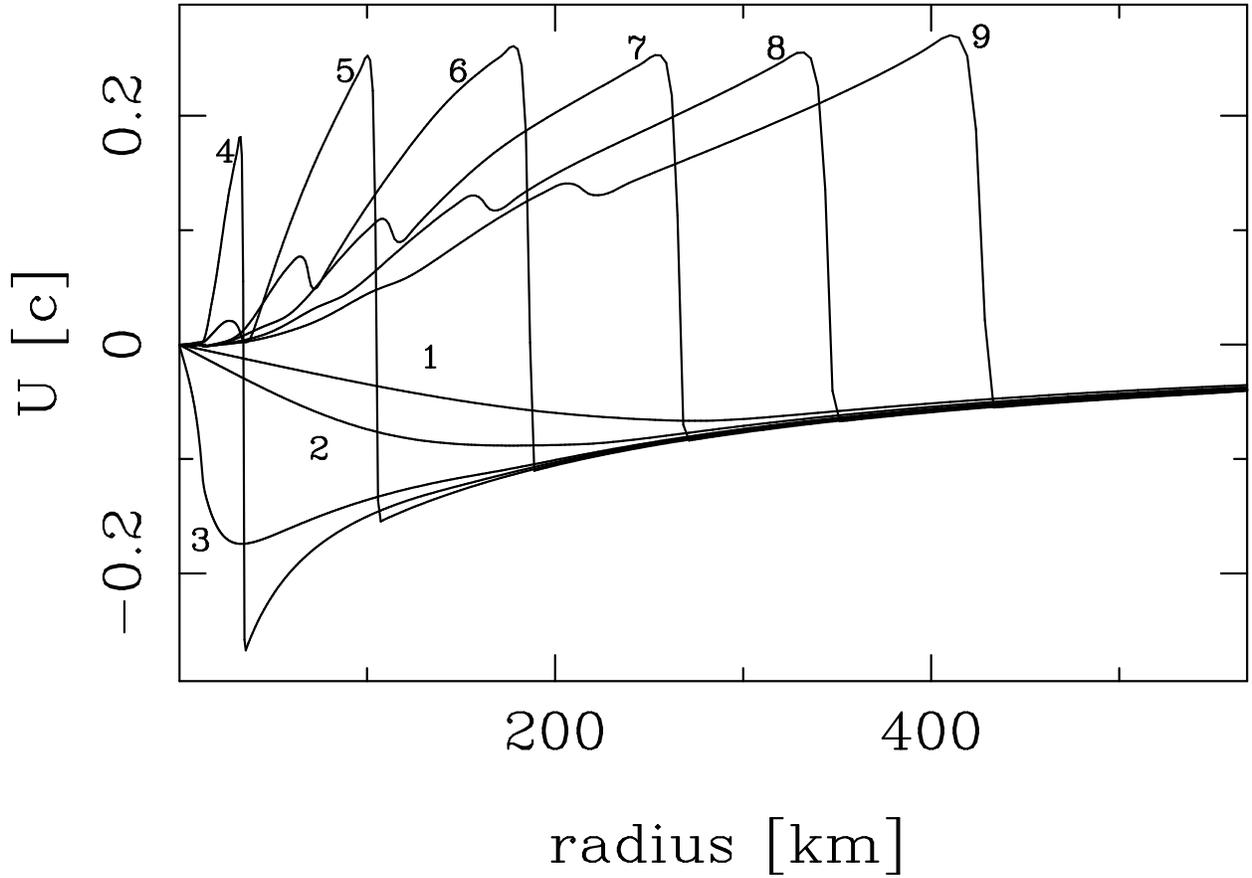} }
\caption[] {Snapshots of the fluid velocity $U$, as a function of the
radius, at different moments of the collapse and bounce, for the
model A of Tab.~\ref{t:efen}. Labels indicate time evolution (1:t=71.2
ms; 2: 76.3 ms; 3: 80.0 ms; 4:81.3 ms; 5: 82.4 ms; 6: 83.5 ms; 7: 84.7
ms; 8: 85.8 ms; 9: 86.9 ms).}
\label{f:gva1}
\end{figure}

\begin{figure}
\centerline{ \epsfig{figure=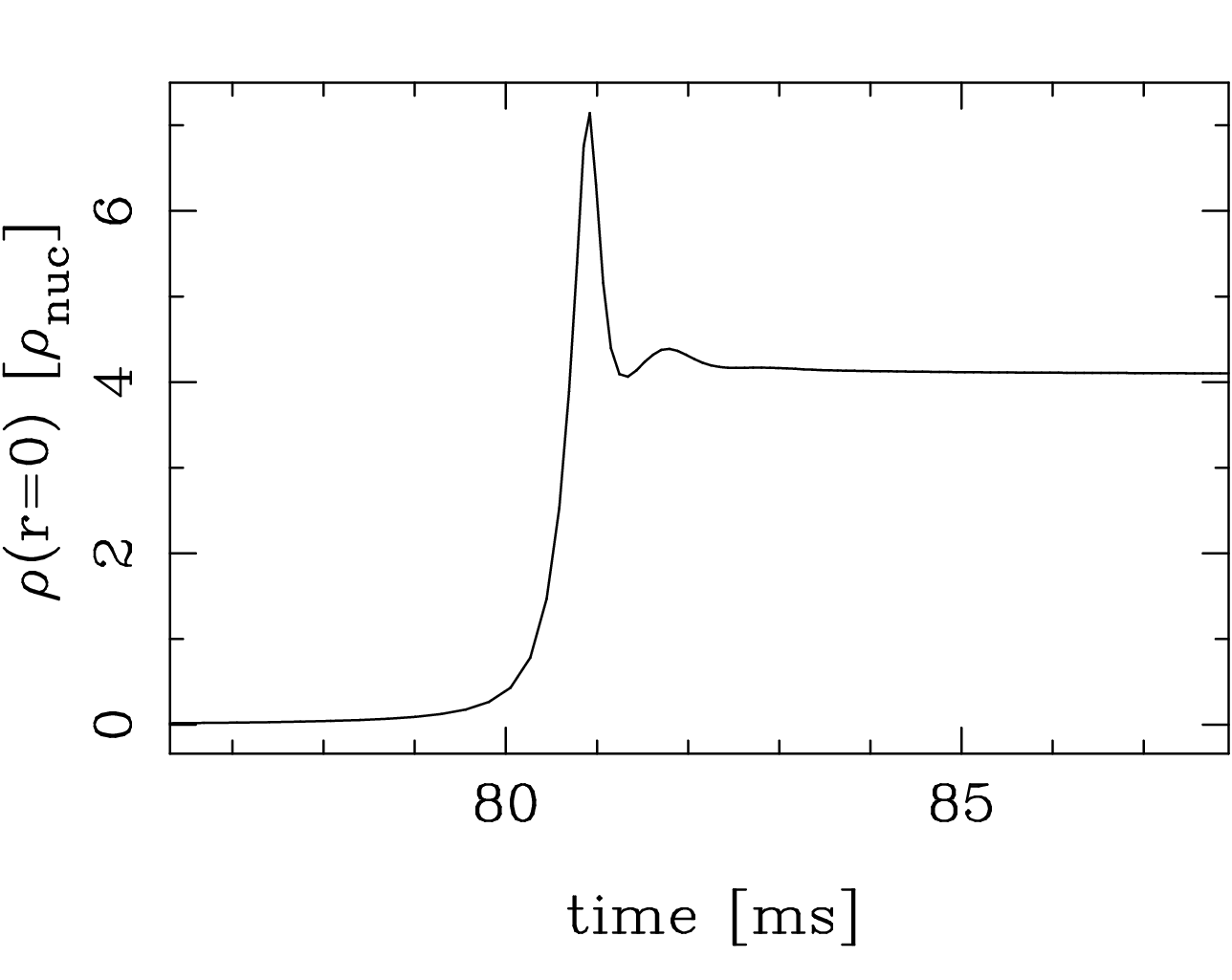,height=\hsize,angle=-90} }
\caption[] {Evolution in time of the central density of the star, for
the model A of Tab.~\ref{t:efen}}
\label{f:rhoA1}
\end{figure}

\begin{figure}
\centerline{ \epsfig{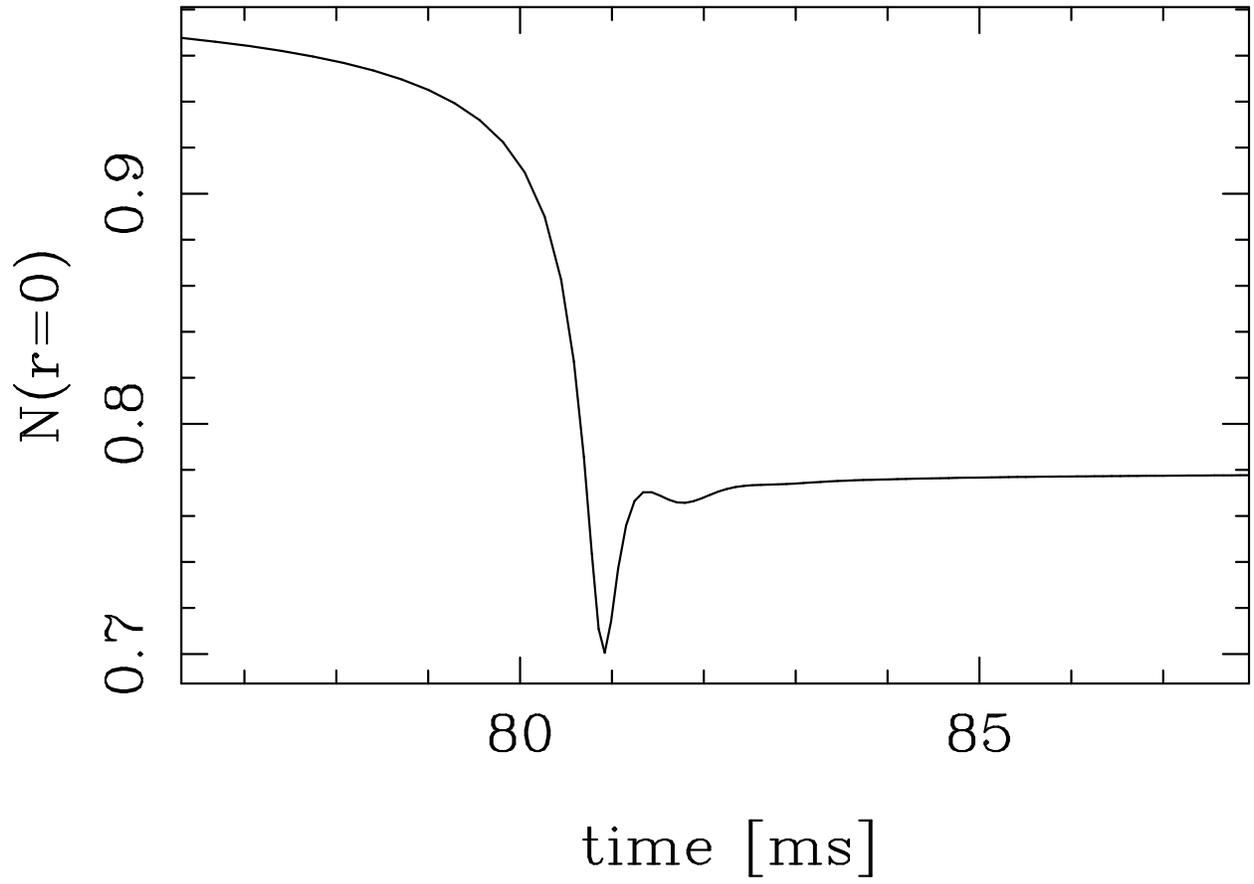} }
\caption[] {Evolution in time of the lapse at the center of the star, for
the model A of Tab.~\ref{t:efen}}
\label{f:lapsA1}
\end{figure}

\begin{figure}
\centerline{ \epsfig{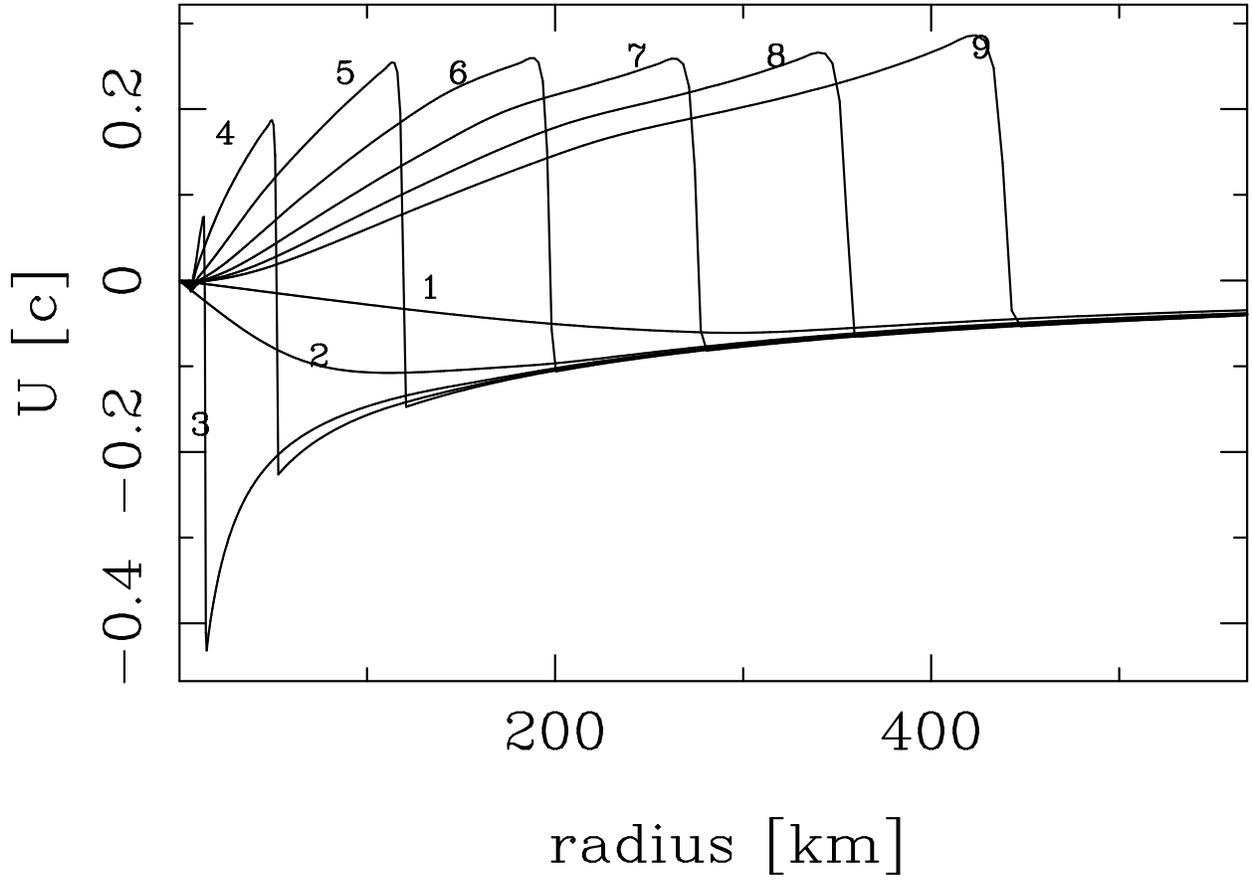} }
\caption[] {Snapshots of the fluid velocity $U$, as a function of the
radius, at different moments of the collapse and bounce, for the
model B of Tab.~\ref{t:efen}. Labels indicate time evolution(1:t=73.8
ms; 2: 79.6 ms; 3: 81.1 ms; 4:81.8 ms; 5: 82.6 ms; 6: 83.4 ms; 7: 84.2
ms; 8: 85.1 ms; 9: 85.9 ms).}
\label{f:gva2}
\end{figure}

\begin{figure}
\centerline{ \epsfig{figure=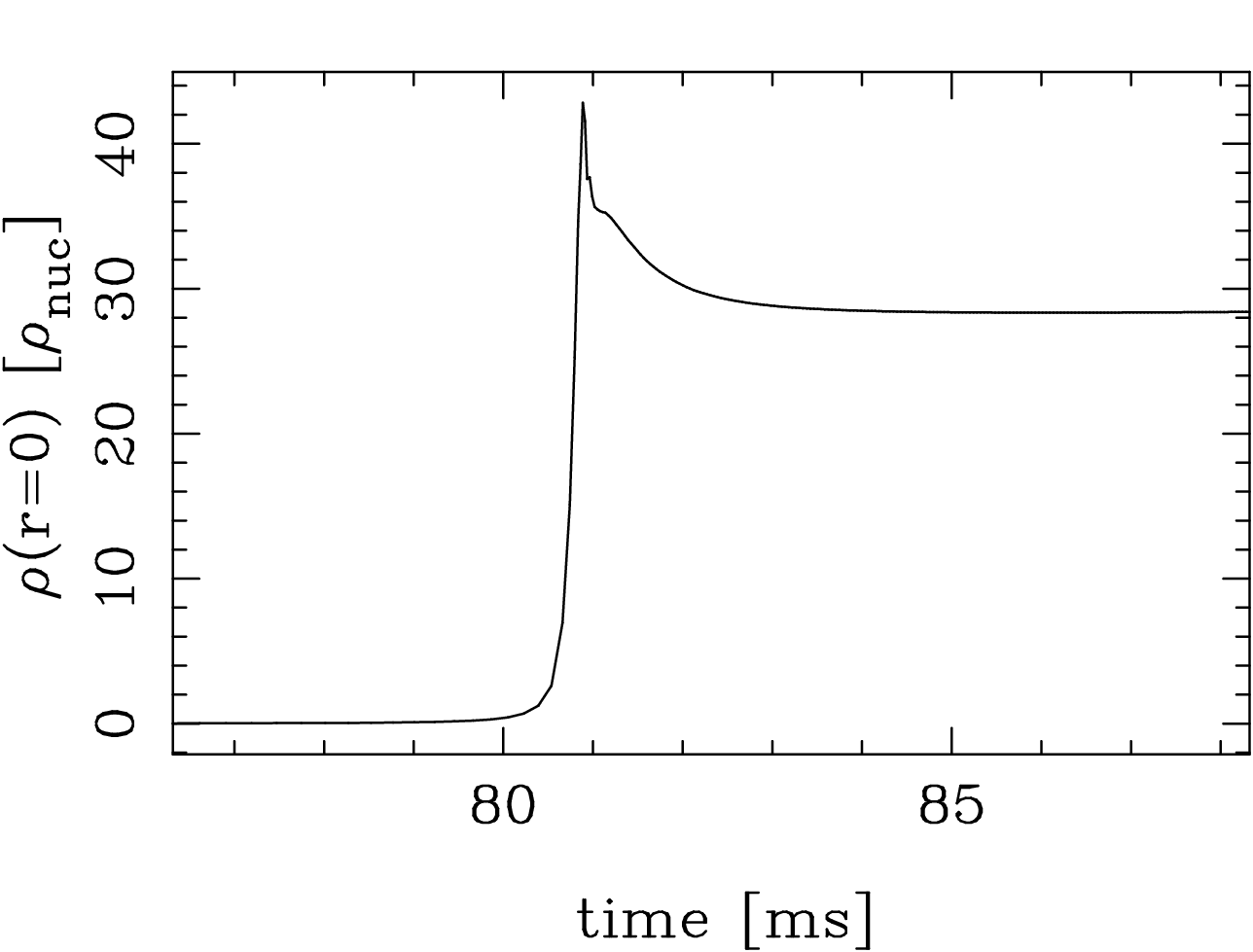,height=\hsize,angle=-90} }
\caption[] {Evolution in time of the central density of the star, for
the model B of Tab.~\ref{t:efen}}
\label{f:rhoA2}
\end{figure}

\begin{figure}
\centerline{ \epsfig{figure=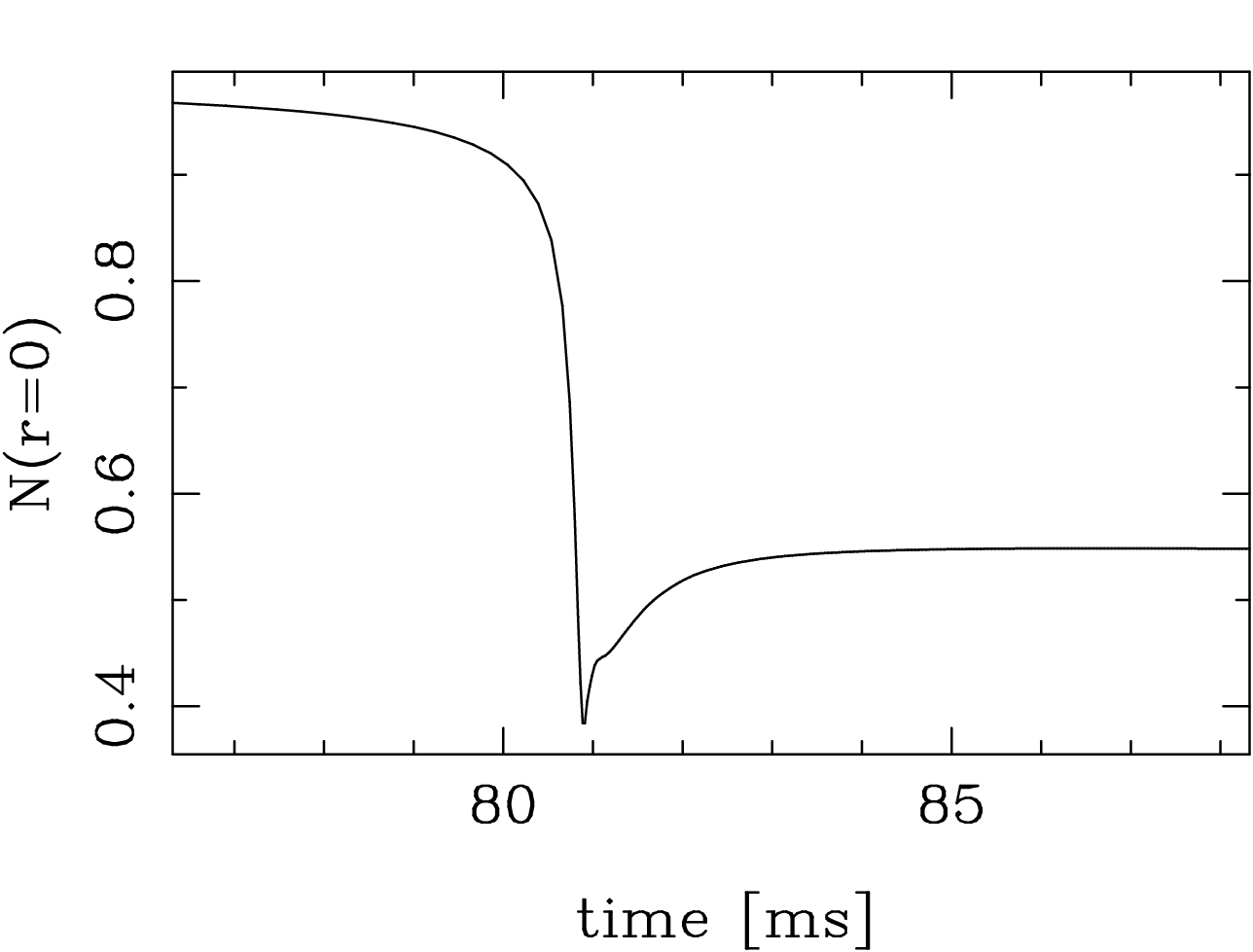,height=\hsize,angle=-90} }
\caption[] {Evolution in time of the lapse at the center of the star, for
the model B of Tab.~\ref{t:efen}}
\label{f:lapsA2}
\end{figure}

We briefly show the results of general-relativistic simulations
(models A and B of Tab.~\ref{t:efen}). We retrieve the results of
\cite{RIM96}, but with the combined code, which is evolving a constant
(equal to $\varphi_0$) 
scalar field (with a null coupling function
(\ref{e:cfunc})). Figs.~\ref{f:gva1} and \ref{f:gva2} gives the
velocity profiles for both simulations: both show the characteristic
``V'' shape during the infall epoch (labels 1,2 and 3 for
Fig.~\ref{f:gva1}, and 1 and 2 for Fig.~\ref{f:gva2}). When the
density near the center becomes larger than $\tilde{\rho}_{bounce}$
(see Figs.~\ref{f:rhoA1} and \ref{f:rhoA2}),
the infall is stopped and the matter coming from outer layers is
stopped and a shock appears propagating outward (labels 4 to 9 of
Fig.~\ref{f:gva1} and 3 to 9 of Fig.~\ref{f:gva2}).
Central density reaches its maximum when the bounce happens
(Figs.~\ref{f:rhoA1} and \ref{f:rhoA2}) and then settles to an
equilibrium value. The evolution of the lapse is similar (see
Figs.~\ref{f:lapsA1} and \ref{f:lapsA2}), and one can see that the
final state of model B is more compact than A, even if the value of
the central density is not compatible with these which are usually
admitted for neutron stars (see e.g.~\cite{SBG94}).

\begin{figure}[p]
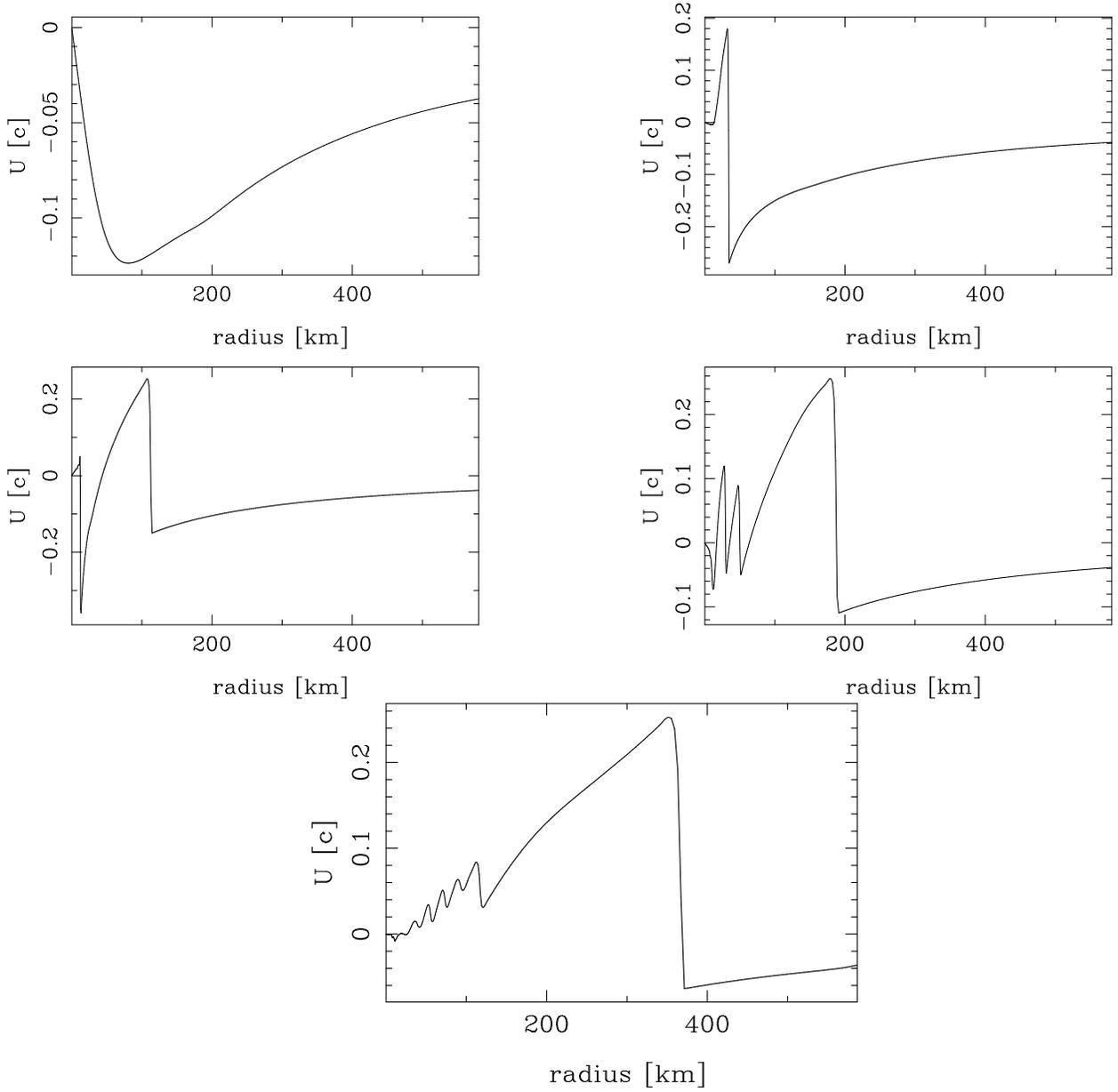

\begin{minipage}[t]{0.45\linewidth}
\hspace{-0.5cm}
\epsfig{figure=gv1c1.ps,width=0.7\linewidth,angle=-90}
\end{minipage}
\hspace{0.5cm}
\begin{minipage}[t]{0.45\linewidth}
\epsfig{figure=gv2c1.ps,width=0.7\linewidth,angle=-90}
\end{minipage}
\begin{minipage}[t]{0.45\linewidth}
\hspace{-0.5cm}
\epsfig{figure=gv3c1.ps,width=0.7\linewidth,angle=-90}
\end{minipage}
\hspace{0.5cm}
\begin{minipage}[t]{0.45\linewidth}
\epsfig{figure=gv4c1.ps,width=0.7\linewidth,angle=-90}
\end{minipage}
\begin{minipage}[t]{\linewidth}
\begin{center}
\epsfig{figure=gv5c1.ps,width=0.35\linewidth,angle=-90}
\caption[] {Profiles of the velocity, as a function of the radius, 
for the model E of Tab.~\ref{t:efen}, before and after the
bounce(from left to right and top to bottom). Times are the following
(t=80.0 ms, 81.3 ms, 82.3 ms, 83.0 ms and 84.6 ms). The 
oscillations behind the shock are due to the oscillation of the
new-born neutron star.}
\label{f:gv5c1}
\end{center}
\end{minipage}
\end{figure}

\begin{figure}
\centerline{ \epsfig{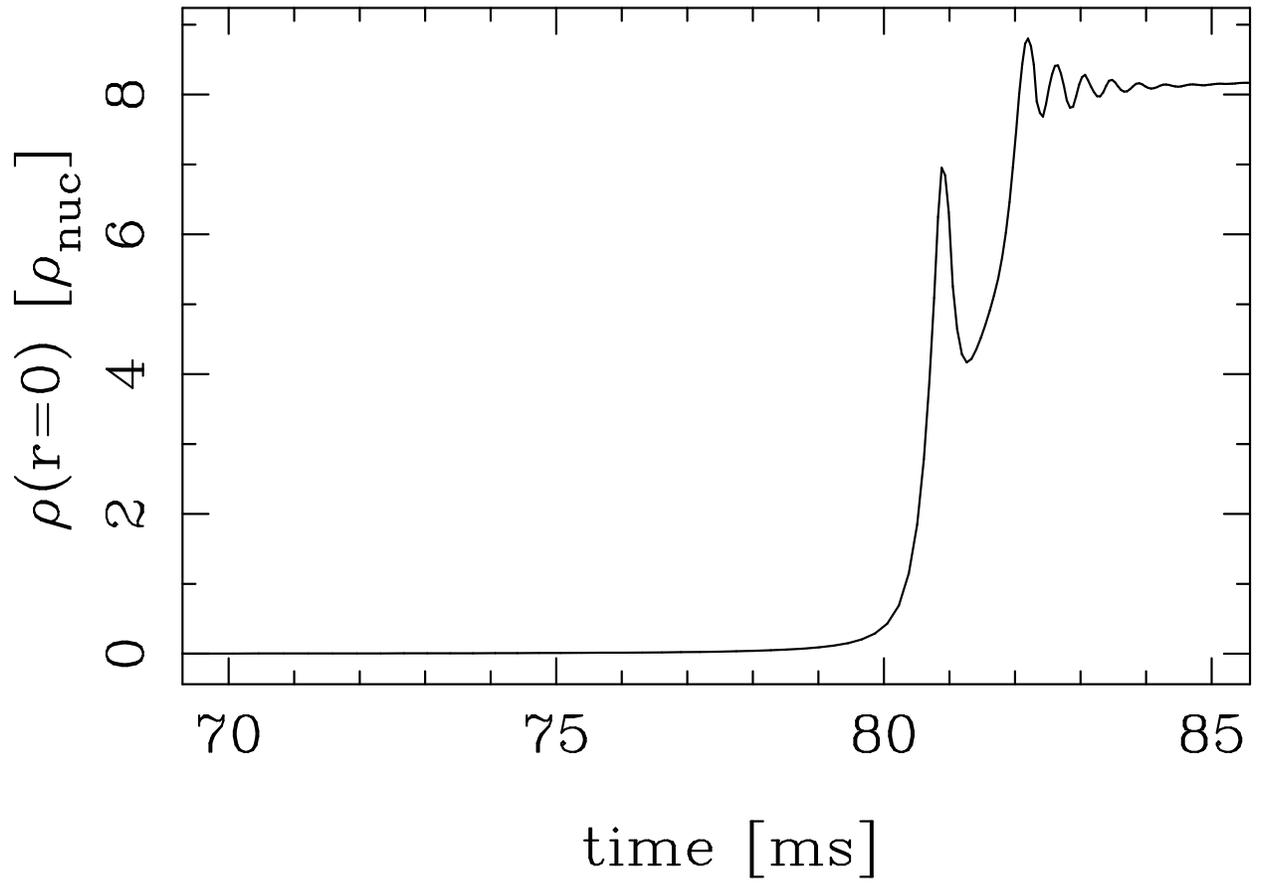} }
\caption[] {Evolution in time of the density at the center of the star, for
the model E of Tab.~\ref{t:efen}}
\label{f:rhoC1}
\end{figure}

\begin{figure}
\centerline{ \epsfig{figure=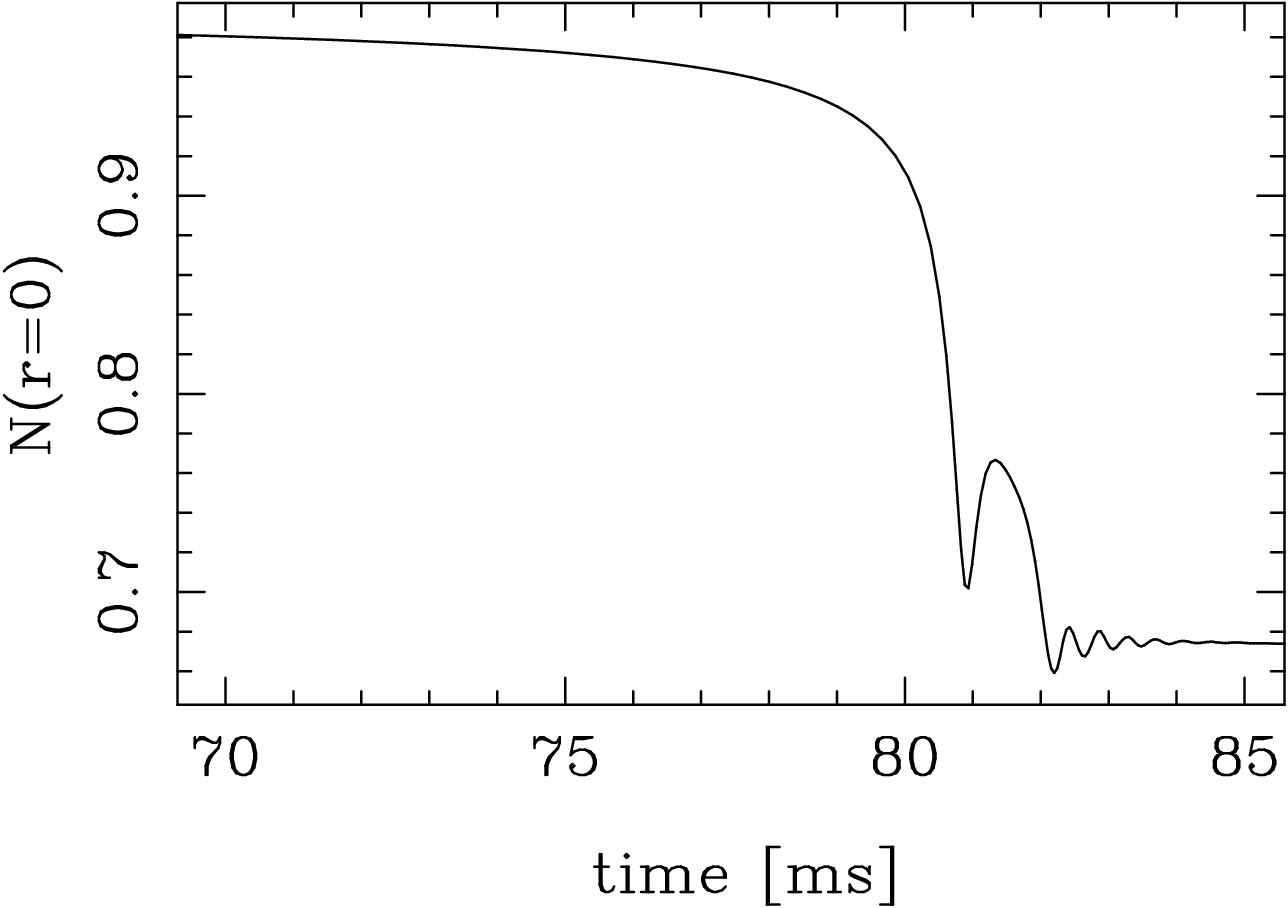,height=\hsize,angle=-90} }
\caption[] {Evolution in time of the lapse at the center of the star, for
the model E of Tab.~\ref{t:efen}}
\label{f:lapsC1}
\end{figure}

\begin{table}
\begin{center}
\begin{tabular}{||c|c|c|c|c|c||}
\hline
model & A & B & C & D & E \\
\hline
 $M_g$ [$M_{\sun}$] & 1.2 & 1.1 & 1.2 & 1.1 & 1.2 \\
 $R_{-3}$ [km] & 31 & 15 & 31 & 15 & 21 \\
 $\omega$ [$M_{\sun}$] & 0 & 0 & $2.3\cdot 10^{-2}$ & $7\cdot 10^{-2}$
& $0.52$ \\
 $E_{\rm{kin}}$ [$10^{-3} M_\odot c^2$] & $2.6$ & $4.1$ & $2.6$ &
 $3.8$ & $1.9$ \\
 $\rho_{\rm{max}}$ [$\rho_{\rm{nuc}}$] & 7 & 43 & 7 & 43 & 8.1 \\
 $N_{\rm{min}}$ & 0.71 & 0.38 & 0.71 & 0.38 & 0.64 \\
 $r\delta \varphi$ [m] & 0 & 0 & 27 & 100 & 800\\
 $E_{\rm{rad}}$ [$10^{-3} M_{\sun} c^2$] & 0 & 0 & $1.3 \cdot 10^{-3}$
& $3.2\cdot 10^{-2}$ & 0.8 \\ 
\hline
\end{tabular}
\caption{\label{t:resen}
Various quantities measured during the numerical evolution of models
described in 
Tab.~\ref{t:efen}.  $M_g$ is the gravitational mass of the resulting
neutron star, $\omega$ its scalar charge (defined as the
monopolar part of the static scalar field, see \cite{DEF93} for
more details) and $R_{-3}$ is such that
$\tilde{\rho}(R_{-3}) = 10^{-3}\tilde{\rho}(r=0)$. $E_{\rm{kin}}$ and
$E_{\rm{rad}}$ are the kinetic energy of the matter reaching the
escape velocity when the shock comes to 
$r=600$ km, and the radiated gravitational energy as a monopolar
wave. $\rho_{\rm{max}}$ is the maximal value reached by the central
density $N_{\rm{min}}$ the minimal value of the lapse
and $r\delta \varphi$ is related to the amplitude of the scalar
gravitational wave by (\ref{e:amplih}). }
\end{center}
\end{table}

We now turn to the tensor-scalar theory with the numerical evolution
of models C, D and
E of Tab.~\ref{t:efen}. Cases C and D have coupling function
parameters that do not allow for the so-called ``spontaneous
scalarization'' (see \cite{NOV98b}), for $\beta_0$ is not negative
enough. Table~\ref{t:resen} summarizes some results of the
simulations. The
most important difference between these runs and the general
relativistic ones comes from the gravitational
radiation emitted during the collapse (see Fig.~\ref{f:fioB1} and
\ref{f:fioB2}), which will be discussed hereafter.
These two figures also show the evolution of the scalar field during
the collapse, the differences between both being related to the fact
that the case D is more relativistic than C. The resulting neutron
star being more compact for D, the scalar field reaches higher values
and its influence on the final state begins to be noticeable. In both
cases, in the beginning (until $t=82$ ms), the scalar field only
``follows'' the matter, producing the spike at $t=82$ ms in both
figures. Then, in the C case, the scalar field varies as the star's
central density. Nevertheless, in the D case, the scalar field has not
reached its equilibrium value yet, so it continues to grow. However,
the differences between  
general-relativistic and weak scalar field runs (A and C or B and D)
are very small, the only one noticeable is the difference in the
kinetic energy of the matter reaching escape velocity, between
runs B and D. This can be seen as if the scalar field deepens the
gravitational potential (which will be confirmed by the study of the
case E), by locally increasing the gravitational coupling function
(see \cite{DEF92}). The figures showing the evolution of central density, of
the lapse and snapshots of velocity distribution are not displayed,
for they are indistinguishable from those of General Relativity.    

The case E has parameters of the coupling function (\ref{e:cfunc})
allowing {\it a priori} for a spontaneous scalarization of the neutron
star. The results shown in Tab.~\ref{t:resen} suggest that has indeed
happened: scalar charge and resulting gravitational radiation are (at
least) one order of magnitude higher than for the C case. The
figure~\ref{f:gv5c1} shows the fluid velocity profile after the
bounce, which exhibits a series of oscillations behind the
shock. These oscillations are no numerical noise since a change in the
number of points in the grids gives the same result and, most of all,
the frequency of these oscillations corresponds to that of a neutron
star undergoing a transition to a ``spontaneous scalarization'' state
(\cite{NOV98b}). It is particularly interesting to compare Fig.~4 in
this reference and Fig.~\ref{f:rhoC1} of this work. Let us recall that
the numerical code used in \cite{NOV98b} is using only {\it spectral
methods} and has been extensively tested; moreover, the particular
simulation of transition to ``spontaneous scalarization'' has been
verified by comparing the final data of the evolution (which represent
a stable equilibrium configuration) with the results given by a static
code. As it can be seen of Figs.~\ref{f:rhoC1} and
\ref{f:lapsC1}, the whole proto-neutron star oscillates, looses energy
through monopolar radiation and settles down to a strong scalar field
state. We also have the confirmation that the scalar field deepens the
gravitational potential well: comparing the run E to A and C, one sees
that the neutron star is more compact and that the ejected matter has
less kinetic energy.

\subsection{Gravitational waves} \label{ss:gwave}
 
\begin{figure}
\centerline{ \epsfig{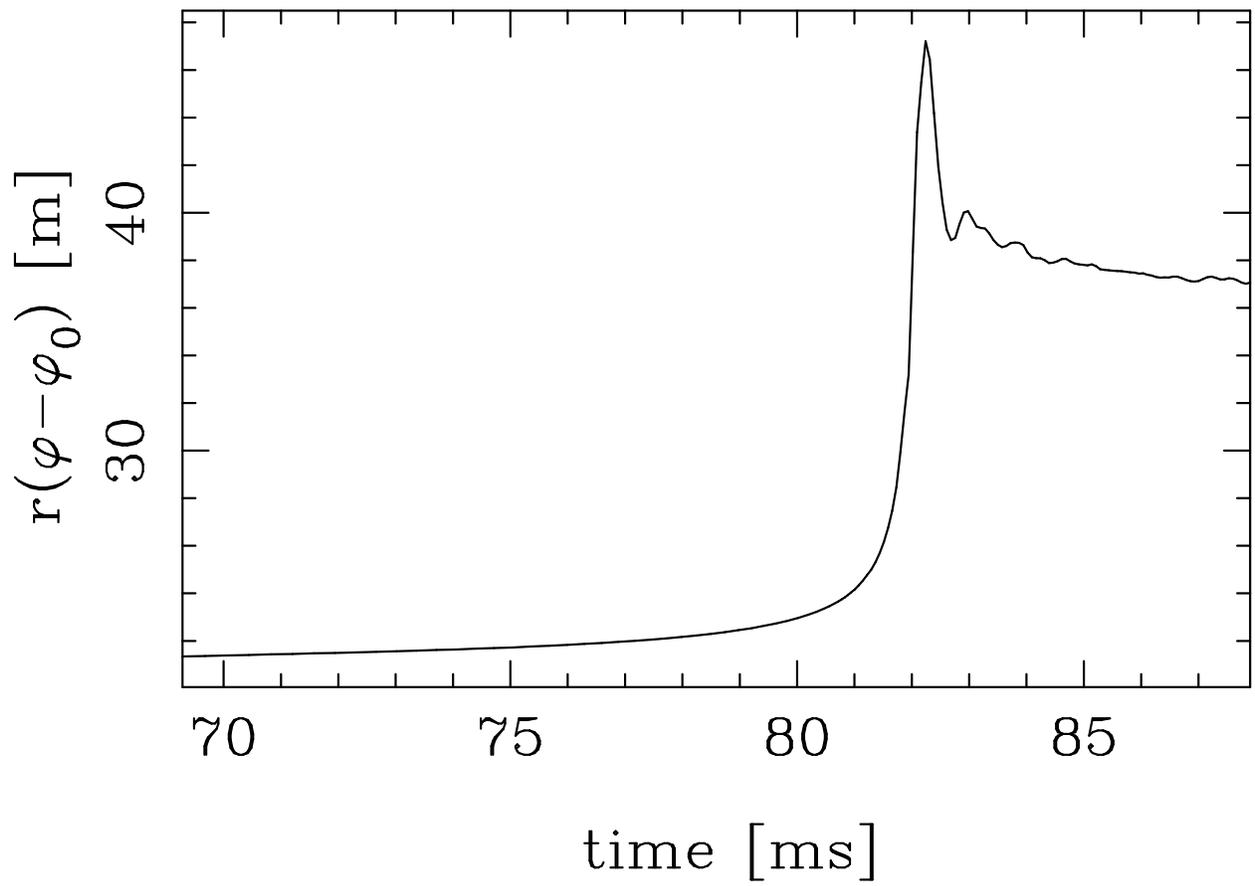} }
\caption[] {Resulting monopolar gravitational waveform (as a function
of time), for the model C of Tab.~\ref{t:efen}}
\label{f:fioB1}
\end{figure}

\begin{figure}
\centerline{ \epsfig{figure=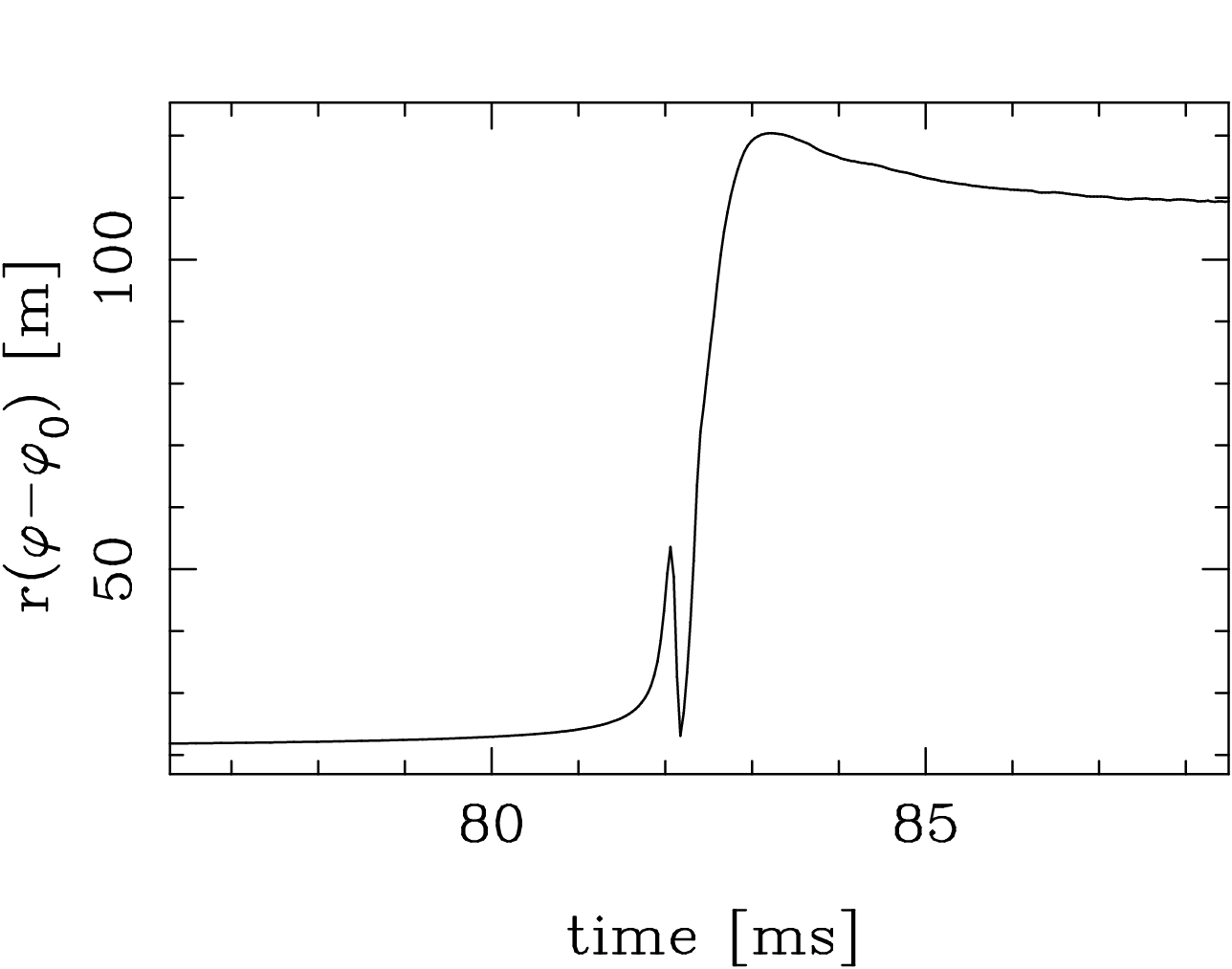,height=\hsize,angle=-90} }
\caption[] {Resulting monopolar gravitational waveform (as a function
of time), for
the model D of Tab.~\ref{t:efen}}
\label{f:fioB2}
\end{figure}

\begin{figure}
\centerline{ \epsfig{figure=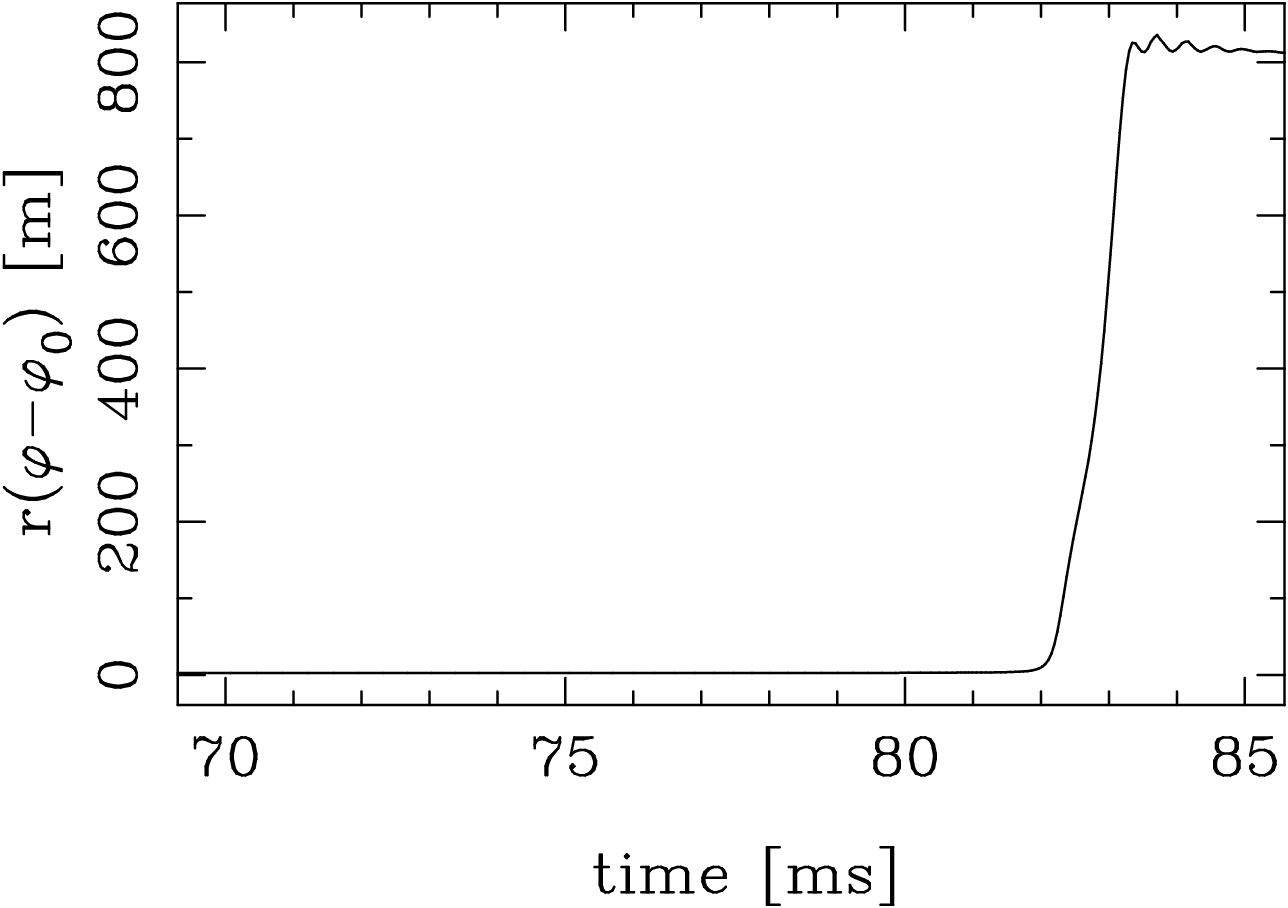,height=\hsize,angle=-90} }
\caption[] {Resulting monopolar gravitational waveform (as a function
of time), for
the model E of Tab.~\ref{t:efen}}
\label{f:fioC1}
\end{figure}

We here recall the results of \cite{DEF92} and \cite{WK97},
concerning the interaction of a scalar wave with interferometric
detectors. Far from any source of gravitational radiation and in a weak
field, the metric writes:
\begin{equation}
\tilde{g}_{\mu\nu} = a^2(\varphi_0) \l( \eta_{\mu\nu} + {1\ov R} \l[
h^*_{\mu\nu} + 2\alpha_0 F \eta_{\mu\nu} \r] + O(1/R^2) \r);
\label{e:metloin2} 
\end{equation}
where $\eta_{\mu\nu}$ is the flat (Minkowskian) metric, $h^*_{\mu\nu}(t-R/c)$
and $F(t-R/c)$ are the $1/R$ components of the Einstein metric
$g^*_{\mu\nu}$ and of the scalar field $\varphi$. The Riemann tensor
of the physical metric can be written (at the first order) as the sum
of two terms:
$$
\tilde{R}^{(2)}_{0i0j} = -{a^2(\varphi_0)\ov 2Rc^2} \dderp{h^{*TT}_{ij}}{t},
$$
which is the spin-2 contribution to the wave ($TT$ denotes the {\it
transverse-traceless} part) and 
$$
\tilde{R}^{(0)}_{0i0j} =  {\alpha_0 a^2(\varphi_0)\ov Rc^2} \l( \delta_{ij}
\dderp{F}{t} - {\partial^2 F \ov \partial x^i \partial x^j} \r).
$$
It is thus clear that monopolar waves interact with the detectors as
well as quadrupolar ones. It is known that if one writes the geodesic
equations in the transverse-traceless gauge, one sees that test
masses have no {\em coordinate} change when a quadrupolar
gravitational wave passes. But the photons measuring the distance
between these two masses are affected by the wave. The opposite holds
for monopolar waves: the photon travel is not affected (since they
follow null geodesics, they cannot differentiate between Einstein and
Fierz metric), but proof masses undergo coordinate change in the $TT$
gauge as the wave passes.

The right quantity to compare with the quadrupolar wave amplitude at a
distance $d$ from the source
$h_Q=h^{TT}_{ij}/d$ appears to be:
\begin{equation}
h_S(t) = {2 \over d} a^2(\varphi_0)\alpha_0 F(t),
\label{e:amplih}
\end{equation}
where $d$ is expressed in meters. This is
the reason why we 
have plotted the function $F(t)$ in Figs.~\ref{f:fioB1}-\ref{f:fioC1},
as in Paper I. These three figures represent the three type of
waveforms we have encountered. In (Fig.~\ref{f:fioB1}) the scalar
field just ``follows'' the hydrodynamical evolution, having almost no
effect on the hydrodynamical evolution. In (Fig.~\ref{f:fioB2}), it
acts a little on this evolution and undergoes raise even after the
bounce occurred. Finally, in (Fig.~\ref{f:fioC1}) the star undergoes a
``spontaneous scalarization'' and the waveform is very close to those
shown in \cite{NOV98b}.
The parameters on which the waveform depends are essentially the
$\beta_0$ (for the spontaneous scalarization to appear) and the final
compactness of the star, which depends on the equation of state used
(\ref{e:gamaro}). The more compact the star, the stronger the
scalar field. 

The results of Paper I, concerning the dependency of the amplitude on
the scalar field parameter $\alpha_0$ were retrieved: in the case of
spontaneous scalarization the wave amplitude (\ref{e:amplih}) goes as
$\alpha_0$, and as $\alpha_0^2$ otherwise. Since this is the amplitude
which interacts with the detectors and since it contains at least one
factor $\alpha_0$ (coming from the interaction of the wave and the
detector, in weak field regime), which is strongly constrained by
solar-system experiments (\cite{DIC94} and \cite{DEF96}), the
result may seem disappointing, if spontaneous scalarization is excluded (as
suggest recent results from binary-pulsar timing, see \cite{DEF98})
and if one compares to the amount of released energy
(Cf.~\ref{t:resen}. However, if we consider the case of the model D: 
taking $\alpha_0=3\times 10^{-2}$, which is the maximal value
compatible with the solar-system experiments, taking the detectability
limit of VIRGO\footnote{at the considered frequency, the sensitivity
of LIGO is comparable} (\cite{VIRGO2}) $3\times 10^{-23}$ Hz$^{-1/2}$ at 800
Hz (which is the characteristic frequency of our events), one sees
that a collapse occurring closer than $10$ kpc from us (half of our
Galaxy) could be detected.

\section{Conclusions} \label{s:conc}
 
We have successfully studied the gravitational collapse leading to a
neutron star, within the framework of a tensor-scalar theory, and
this is the first study of a gravitational collapse,
with the presence of shocks, in tensor-scalar gravity. We have
shown that, if the parameters of the theory are such that the
non-perturbative phenomenon involving the scalar field (the so-called
``spontaneous scalarization'') can be excluded, as seem to indicate
results of \cite{DEF98}, then the scalar field has a little influence
on the hydrodynamical evolution of the collapse. If the case where
spontaneous scalarization can occur, the run indeed indicated that the
resulting neutron star had a very strong scalar charge and that it
underwent strong characteristic oscillations. This gives us high
confidence in our combined code. As far as gravitational
waves are concerned, we have shown that, with $\alpha_0$ and $\beta_0$
constrained with solar-system tests and binary-pulsar timing, such a
collapse could be detected up to $10$ kpc from us. From the opposite
point of view, a non-detection of any monopolar wave, when a supernova
occurs (when we have neutrino and/or electromagnetic signal), could
give constraints on the parameter space of the tensor-scalar
theory. The computed waveforms appear to be quite different from those
of Paper I, meaning that, if monopolar waves from gravitational
collapses could be detected, one could in principle differentiate
between each type of collapse: Type II Supernova or collapsing neutron
star. As a byproduct, we have developed and tested a new type of
code, combining spectral methods and Godunov-type
techniques. We hope to proceed to more complicated problems (in more
than one dimension) in General Relativity (that is, without the scalar
field), using, at each time step, HRSC methods for solving numerically the 
equations governing the evolution of matter (the Hydro-part) and spectral 
methods for solving numerically the equations governing the gravitational 
field equations (Einstein-part) just for the distribution of matter given 
at that time step by the Hydro-part. The experience gained in present work 
seems to be stimulating for such new strategy in Numerical Relativity.

\acknowledgements

We are indebted to Silvano Bonazzola for providing us with the
``smooth'' interpolation procedure, as well as for very useful
discussions and reading of the manuscript.
JN acknowledges financial support connected with a post-doctoral
fellowship LAVOISIER from the French ministry of foreign affairs. 
This work has been partially supported by the Spanish DGES
(grant number PB97-1432).

\end{document}